\documentclass{appolb}
\usepackage{epsfig}

\def\runhead{}

\newcommand{\nn}{{\cal N}}
\newcommand{\eq}{\begin{equation}}
\newcommand{\eqx}{\end{equation}}
\newcommand{\eqn}{\begin{eqnarray}}
\newcommand{\eqnx}{\end{eqnarray}}
\newcommand{\f}[2]{\frac{#1}{#2}}
\newcommand{\lm}{\lambda}

\newcommand{\lra}{\longrightarrow}

\renewcommand{\th}{\theta}
\newcommand{\sg}{\sigma}
\newcommand{\al}{\alpha}
\newcommand{\alef}{\alpha'_{eff}}
\newcommand{\OM}{\Omega}
\newcommand{\dl}{\delta}

\newcommand{\DD}{{\cal D}}
\newcommand{\AAA}{{\cal A}}

\newcommand{\qb}{\bar{q}}
\newcommand{\cor}[1]{\left\langle{#1}\right\rangle}
\newcommand{\qqqq}{\quad\quad\quad\quad}

\newcommand{\rr}[4]{#1, {\it #2 \/}{\bf #3} #4}

\title{High energy scattering and AdS/CFT}

\author{Romuald A. Janik
\thanks{e-mail:{\tt janik@nbi.dk}.
Talk presented at the XLI Cracow School of Theoretical Physics, 2-11
June 2001.}
\address{The Niels Bohr Institute,\\
Blegdamsvej 17, DK-2100 Copenhagen,\\ 
Denmark\\
and\\
M. Smoluchowski Institute of Physics,\\ 
Jagellonian University,\\
Reymonta 4, 30-059 Cracow, \\
Poland}
}

\begin{document}

\maketitle

\begin{abstract}
In this talk we describe the application of the AdS/CFT correspondence
for a confining background to the study of high energy scattering amplitudes
in gauge theory. We relate the energy behaviour of scattering amplitudes to
properties of minimal surfaces of the helicoidal type. We describe the
results of hep-th/0003059 and hep-th/0010069 for amplitudes with vacuum
quantum number exchange and, very briefly, hep-th/0110024  on the
extension of this formalism to Reggeon exchange. 
\end{abstract}

\section{Introduction}

It is well known that high energy scattering amplitudes with small
momentum transfer (more precisely $t/s \to 0$) are {\em
phenomenologically} well described by the exchange of Regge poles (see
e.g. \cite{la92}). The
dominant contribution to scattering processes with no exchange of
quantum numbers is the exchange of the Pomeron. These amplitudes
behave like $s^{1.08+0.25t}$. Other processes, involving the exchange
of a Reggeon, have a different behaviour of the type $s^{0.5+1 t}$.
The theoretical description of such proccesses from first principles
remains a formidable 
challenge, as it is of an inherently nonperturbative character. 
In this talk we will describe an approach developed in
\cite{us1,us2,fluct,usr,SRZ} which uses the
AdS/CFT correspondence. 

The AdS/CFT correspondence \cite{ma98,ma99} was first proposed as an
equivalence between $\nn=4$ supersymmetric gauge theory and type IIB
string theory in a curved $AdS_5\times S^5$ background. The statement
that the two 
theories are indeed equivalent means that all (gauge-invariant)
observables in the SYM should be calculable in the string theory
language. At least for some observables like Wilson loops such
prescriptions are 
available (but there is no direct proof up till now).
The utility of this correspondence for nonperturbative calculations in
gauge theory comes from the fact that calculations done for large
't-Hooft coupling ($\lm=g^2_{YM}N$) are translated into
(semi-)classical calculations on the string theory side.

The original version of the AdS/CFT correspondence involved a
conformal supersymmetric gauge
theory, but later it was generalized to other cases including
confining theories. We will use one such version to study the interplay
between confinement and properties of soft high energy scattering
amplitudes.
Our results should be, however, quite generic.

 
Finally let us mention that other approaches to the description of the
nonperturbative scattering physics exist and use various models
of the nonperturbative vaccuum like the instanton vaccuum \cite{SZN,KKL} and
Stochastic Vacuum Model \cite{Nacht,Nachtr}.
  
The outline of this talk is as follows. In section 2 we will describe
the basic features of the AdS/CFT correpondence, in section 3 we will
state the appropriate formulation of scattering amplitudes in gauge
theories which is a convenient starting point for performing the
calculation using the AdS/CFT correspondence. Then we will move 
on to evaluate the classical contribution to the scattering amplitude
which is determined, in this framework, by solving a minimal surface
problem. In section 6 we will evaluate the contribution of quadratic
fluctuations and show how it gives rise to a shift of the intercept.
In section 7 we give a brief discussion of Reggeon exchange. 
We close the paper with a summary and outlook.

\section{The AdS/CFT correspondence}

The AdS/CFT correspondence was discovered by Maldacena \cite{ma98}, who
conjectured that two seemingly unrelated theories, namely  $\nn=4$ SYM gauge
theory and type IIB string theory in an $AdS_5\times S^5$ space are in
fact completely equivalent. This statement gives a very concrete
realization of the old hope that gauge theories can be described by
string theory in the large $N$ limit. The novel feature is the fact
that strings of the dual string theory live in 10 dimensions, the
geometry of which is 
moreover curved. These extra dimensions could be roughly understood as
some parameter space for some (unspecified at the moment)
semiclassical field configurations in gauge theory. An example of such
an interpretation is the identification of instantons in $\nn=4$ SYM
with $D(-1)$ string instantons (which are points in $AdS_5 \times
S^5$). The $5^{th}$ coordinate of these points represents  the size
of the instanton
in $\nn=4$ SYM \cite{BGKR}. So the string theory side of the AdS/CFT
correspondence can be thought 
of to be a convenient and very nontrivial
choice of degrees of freedom in the original gauge theory especially
suited to studying properties at large gauge coupling (see below).

A difference with the old `effective string' picture of QCD is
that this correspondence is thought to be exact and not only valid in
the IR. The various properties of different gauge theories are encoded
in differing background geometries for the string theory. 

The utility of the AdS/CFT correspondence lies in the fact that when
the gauge coupling ($g^2_{YM}N$) is large, gravity (in the 10D string
background geometry --- this has nothing to do with real world
gravity!) becomes weak
and (semi-)classical methods in string theory become applicable.

As it was already mentioned, the correspondence was later extended to
different gauge theories, some of which were confining. Although an
exact counterpart for QCD is unknown, we will perform the calculations
in a specific confining background and argue that the main features of
our approximation scheme are generic and do not depend on the very
detailed properties of the background geometry.    

Before we proceed let us briefly recall the way in which the AdS/CFT
correspondence arose. The idea is to construct a gauge theory from
string theory and to look at this same construction from two
perspectives, each of which can be thought 
of to
give a different picture of the same underlying theory. These two
points of view become just the two sides of the AdS/CFT correspondence.

The first path to gauge theory is to consider $N$ coincident D3 branes in flat
10D space. The D3 branes are nonperturbative objects in type IIB string
theory. They are 
4-dimensional hypersurfaces on which open strings may
end. The massless excitations of these open strings form the
multiplet of $\nn=4$ gauge theory. In addition there are massive
excitations with the mass scale set by $m^2 \propto 1/\al'$. When one
takes the $\al' \to 0$ limit, these additional states will decouple
and one will be left with just the gauge theory (and a noninteracting
flat 10D bulk theory).

Let us now take into account that the D3 branes are massive
and charged objects. Therefore a stack of $N$ D3 branes will curve
spacetime and generate some metric for the closed strings. 
After an appropriate rescaling of
coordinates one then performs the same $\al' \to 0$ limit as
before. The result is the $AdS_5 \times S^5$ background for closed
strings. Since this should be exactly the same system as before one is
led to the AdS/CFT correspondence.

\subsection*{Wilson loops}

As an example of using the AdS/CFT correspondence let us recall the
prescription for calculating the expectation values of Wilson
loops. We will use it later for evaluating scattering amplitudes.

The generic feature of the 10D dual geometries is that they posses a
well defined boundary. The prescription for calculating $\cor{W(C)}$
is to place the contour $C$ on the boundary, and consider
the partition function of strings in the AdS bulk which are spanned on
C \cite{loopsads}:
\eq
\cor{W(C)} \sim \int_{\partial \Sigma=C} {\DD X^A} e^{-\f{1}{2\pi
\al'} S_{string}(\Sigma)} 
\eqx 
where $\Sigma$ is the string worldsheet defined by the bosonic fields
$X^A$, $A=0..9$ (here we suppressed the fermionic partners), $S_{string}$
is the string action. $1/\al'$ typically involves $\sqrt{g^2_{YM}N}$,
so at large gauge coupling the partition function is saturated by the
string worldsheet spanned on $C$ whose surface is minimal:
\eq
\cor{W(C)} \sim Fluctuations(C) \cdot e^{-\f{1}{2\pi \al'}
Area(\Sigma_{minimal})} 
\eqx
We will now illustrate this prescription with a calculation of the
static $q\qb$ potential in the case of (a) the conformal $\nn=4$
theory and (b) a confining theory.

\subsubsection*{$\nn=4$ SYM theory}

In this case the background geometry is pure $AdS_5 \times S^5$:
\eq
ds^2=\f{dz^2}{z^2}+\f{dx^\mu dx^\mu}{z^2}+d\OM_5
\eqx 
The boundary is at $z=0$. The problem of calculating the potential
boils down to finding $\cor{W(T\times R)}$ with $T\to \infty$. Thus we
have to put the two lines separated by $R$ on the boundary at
$z=0$. Due to the factors of $1/z^2$, we see that in order to minimize
distances (and hence the area), it pays to increase $z$ and go into
the bulk, as far from the boundary as possible.
The resulting minimal surface in the bulk was found in \cite{loopsads} and
the (regularized) area was evaluated to yield:
\eq
\cor{W(T\times R)} \sim e^{\f{4\pi^2 \sqrt{2 g^2_{YM}
N}}{\Gamma(1/4)^2} \f{T}{R} }
\eqx  
where we expressed all string-theoretic parameters in terms of the
gauge coupling. In these calculations one always subtracts a
divergence due to singularity of the metric near the boundary
\cite{loopsads,DG}. 

We note that despite the fact that we have a string picture of a
Wilson loop we
obtained a {\em coulombic} potential. This is because of the
nontrivial curved geometry of the string background.

\subsubsection*{Confining black hole background}

\begin{figure}
\centerline{\epsfysize=5cm  \epsfbox{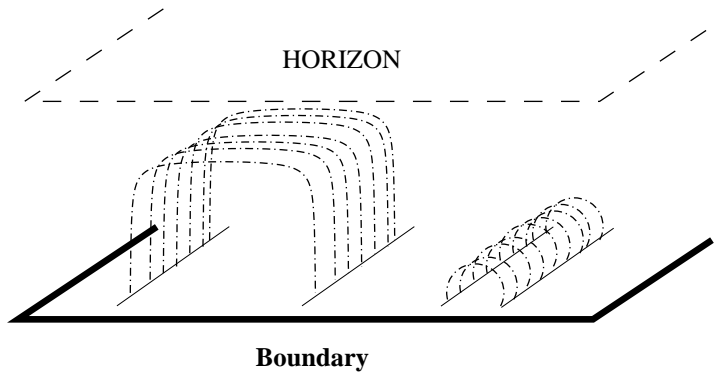}}
\caption{Static $q\qb$ potential using the black hole geometry.}
\end{figure}

The first and simplest background for a confining theory was proposed
by Witten in \cite{wittenbh}. The metric has the form:
\eq
ds^2=\f{16}{9} \f{1}{z^{2/3}(1-(z/R_0)^4)} \f{dz^2}{z^2}+
\f{\eta_{\mu\nu}dx^\mu dx^\nu}{z^2}+ \ldots
\eqx
where $R_0$ is the scale of the horizon and $\ldots$ represent
additional dimensions which do not play any role for the problem at
hand.
Again it pays to increase $z$ as much as possible, but now the range
of possible $z$ is limited by the horizon radius $R_0$. We therefore
have two qualitatively different regimes. When the size of the Wilson
loop $W(C)\equiv W(T\times R)$ is greater than $R_0$, the 
minimal surface will rise up to the
horizon and then will extend at a fixed value of $z_{fixed}$ close to
$R_0$ (see fig. 1). The area of this minimal surface will be
proportional to the area of $C$, since it is measured in the
effectively flat metric 
\eq
ds^2=\f{1}{z_{fixed}^2} dx^\mu dx^\mu
\eqx
Therefore we have
\eq
\cor{W(T\times R)}\sim e^{-\f{1}{2\pi \alef} TR}
\eqx
where we absorbed all the scale factors in the effective string
tension. We see that in this regime the details of the metric
(powers/coefficients) do not matter and we expect a large degree of
universality.  

On the other hand when the size of $C$ is small compared to $R_0$, the
minimal surface problem will depend very much on the structure of the
metric for small $z$ (cf. the smaller loop in fig. 1), and therefore
it will depend on the detailed nature of the gauge theory.

\subsubsection*{Fluctuations}

In the above examples we included just the action of the classical
solution of the equations of motion for the string (the minimal
surface equations). It is often interesting to go further, and
calculate the contribution of quadratic fluctuations around this
solution. In general this is a very difficult problem, so we will now
state the result for the confining black hole background in the
flat space approximation.
The answer in flat space is \cite{luscher,alvarez,frts,arvis}
\eq
\label{e.fltr}
Fluctuations(T\times R) \sim \left(\det
\Delta\right)^{-\f{n_\perp}{2}}=
e^{n_\perp \cdot \f{\pi}{24} \cdot \f{T}{R}}
\eqx 
where $\det \Delta$ is the determinant of the laplacian with Dirichlet
boundary conditions, and $n_\perp$ is the effective number of massless
transverse degrees of freedom for the string. For the superstring in
the black hole background the expected number is $n_\perp=7$ \cite{kinar} (see
section 6). The contribution (\ref{e.fltr}) is called the L\"uscher
term and gives a Coulombic correction to the linear confining
potential. It is independent of the string tension, and in fact quite
universal, in the sense of being independent of most of the details of
the specific string theory.  

\section{Gauge-theoretic scattering amplitudes}

In this section we will rewrite scattering amplitudes in gauge theory
in a form adapted to calculation using the techniques of the
AdS/CFT correspondence.

\begin{figure}[h]
\centerline{\epsfysize=6cm \epsfxsize=5cm \epsfbox{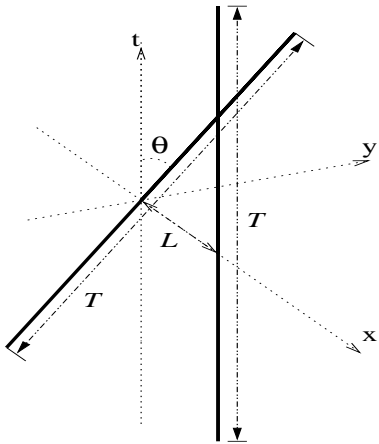}}
\caption{Geometry of the Wilson lines in Euclidean space.}
\end{figure}

It is convenient to consider scattering amplitudes in impact parameter
space which are defined through
\eq
A(s,t)=\f{is}{2\pi} \int d^2l\, e^{iq\cdot l}\, \tilde{A}(s,L=|l|)
\eqx 
Since we want to study soft proccesses with small momentum transfer
(and no exchange of quantum numbers), we may use the eikonal
approximation \cite{Nachtr}. In this approximation the impact
parameter $q\qb$ scattering 
amplitude is given by a correlation function of two Wilson lines which
follow classical straight line trajectories (see fig. 2):
\eq
\label{e.lines}
A(s,L)=\cor{e^{i\int_{L_1} A} e^{i\int_{L_2} A}}
\eqx
This expression suffers from two drawbacks. It is IR divergent and it
is not gauge invariant. The first property requires us to work with an
IR cut off i.e. a finite temporal length $T$ of the lines. In addition
we also have to add a gauge connector between the end points in order
to obtain a gauge invariant quantity. Alternatively we may choose to
work with a gauge invariant and IR finite amplitude for the scattering
of two $q\qb$ {\em pairs}. Then the lines in (\ref{e.lines}) are
replaced by Wilson loops, closed at infinity by mesonic wave
functions.

The latter quantity has a very clear physical significance, however,
it is more difficult to calculate. Therefore we will
concentrate on the $q\qb$ scattering with an explicit IR cut-off. In
the context of a confining theory which we will consider, such an IR
cut-off may have a direct physical significance as a scale on which colour
strings start to break. Another argument in favour of studying this
simpler situation is that a variational treatment of the scattering of
two $q\qb$ pairs in the same framework \cite{us2}, essentially leads to a
superposition of $q\qb$ scatterings with small IR cut-off $T$.  

It will turn out to be convenient to perform one more reformulation of
the problem and to go to {\em Euclidean} signature. There we will
calculate the correlation function of the lines as a function of the
relative angle $\th$, staying within Euclidean gauge theory. The
result $A(\th,L)$ will then be continued back to {\em Minkowski} space
using the substitutions
\eq
\th \lra -i\chi \sim -i \log s \qqqq T\to i T
\eqx
The above procedure was first used within the eikonal approximation in
perturbative QED and QCD in \cite{Megg}.

Since the scattering amplitude is expressed as a correlation function
of two Wilson lines, we may apply now the methods of the AdS/CFT
correspondence in a confining background to calculate it.

\section{The evaluation of scattering amplitudes}

Let us now evaluate the correlation function of two Wilson lines
inclined at an angle $\th$ and separated by a transverse distance $L$.
To this end, following the prescription of section 2, we have to
place the two lines on the boundary and find the minimal surface in the
black hole geometry. Again two regimes will appear, depending on the
relation of the impact parameter to the confinement scale $R_0$. When
the impact parameter is sufficiently large the same arguments should
apply as for the study of the confining potential: the minimal surface
problem is essentially transported up to the horizon $z\sim R_0$,
where the minimization occurs in flat space. 
This is the case that we are going to study.
We thus have to find a
minimal surface in {\em flat space} between two lines at an angle.

\section{Minimal surface}

The appropriate minimal surface is indeed well known -- it is the
helicoid. It can be parametrized by
\eqn
\label{e.helt}
t &=& \tau \cos p\sg\\
y &=& \tau \sin p\sg \\
\label{e.helx}
x &=& \sg
\eqnx
where $\tau=-T \ldots T$, $\sg=-L/2 \ldots L/2$ and $p=\th/L$.
According to the AdS/CFT prescription the correlation function is
equal to 
\eq
e^{-\f{1}{2\pi \alef} Area}=e^{-\f{1}{2\pi \alef} \int d\sg d\tau
\sqrt{\det g_{ab}}} 
\eqx
where $g_{ab}=\partial_a X^\mu \partial_b X^\mu$ is the induced
metric. The explicit formula for the area is \cite{us2}
\eqn
\label{e.area}
Area&=&\int d\sg d\tau \sqrt{\det g_{ab}}=\int_{-L/2}^{L/2} d\sg
\int_{-T}^T d\tau \sqrt{1+p^2\tau^2}=  \\
&=& \int_{-L/2}^{L/2} d\sg 
\, \left\{T \sqrt{1+p^2 T^2}+ \f{1}{p}\log\left( p T +
\sqrt{1+p^2 T^2} \right) \right\}
\eqnx
In order to obtain the scattering amplitude we have to perform
analytical continuation to Minkowski space. A naive continuation would
give a result which is a pure phase. However we see that the formula
for the area involves a logarithm which has a cut in the complex
plane. A priori we cannot rule out the possibility of moving onto a
different sheet of the logarithm when doing the analytical
continuation, especially as we are dealing with an inherently
nonperturbative Euclidean correlation function. 
It is therefore interesting to explore the physical consequences of
moving to a different Riemann sheet. Consequently we have to perform
the substitution 
\eq
\log(\ldots) \lra \log(\ldots)- 2\pi i n
\eqx
with $n$ being some integer number.
Under this transformation, the amplitude gets a contribution:
\eq
e^{\f{1}{2\pi \alef} \f{L^2}{\th} 2\pi i n} \lra 
e^{-\f{1}{\alef} \f{L^2}{\log s} n}
\eqx
which is {\em independent} of the IR cut-off $T$. In the following we
will neglect the $T$ dependent terms assuming that $T$ is small (some
justification for this assumption was given in \cite{us2}). After Fourier
transform we obtain an inelastic amplitude with a linear Regge
trajectory:
\eq
(prefactor) \cdot s^{1+\f{\alef}{4} t}
\eqx
The prefactor here includes a $\log s$, further such contributions may come
from $\al'$ corrections. In the following we concentrate on the
dominant terms which give rise to a power-like $s^\al$ behaviour.

An interesting feature of this result is that the linear slope arose
through the analytic structure of the helicoid area. Moreover the
slope $\alef/4$ characteristic of soft Pomeron exchange appeared in a
natural way\footnote{In the black hole background $\alef$ is a free
parameter, but it is directly linked to the static $q\qb$
potential. Hence we may take the phenomenological value of QCD string
tension as defining $\alef$.}. 

In the next section we will see that quadratic fluctuations of the
string worldsheet around the helicoidal minimal surface give rise to a
shift of the intercept. 

\section{Fluctuations}

In order to perform the calculation of the contribution of the
quadratic fluctuations we have to decompose the fields:
\eq
X^A(\sg,\tau)=X^A_{helicoid}(\sg,\tau)+x^A(\sg,\tau)
\eqx
and expand the string action to second order in $x^A(\sg,\tau)$. 
We assume, as is the case for string theories arising in the AdS/CFT
correspondence, that the action for quadratic fluctuations is just the Polyakov
action. 

In order to simplify the calculation \cite{fluct}, it is convenient to
change the parametrization of the helicoid and replace
the variable $\tau$ in (\ref{e.helt})-(\ref{e.helx}) by
\eq
\rho=\f{1}{p}\log(p\tau + \sqrt{1+p^2 \tau^2}) \ .
\eqx
The advantage of doing this is that the induced metric on the helicoid
w.r.t the variables $\rho$, $\sg$ is
conformally flat i.e. 
\eq
g_{ab}=(\cosh^2 p\rho)\ \dl_{ab} \ .
\eqx
Therefore, since string theory in the AdS background is {\em critical}, we may
perform the calculation for the conformally equivalent  flat metric
$g_{ab}=\dl_{ab}$. The path integral with the Polyakov action gives
\eq
\label{e.detform}
\left( \det \Delta \right)^{-\f{D-2}{2}} \equiv \left( \det \Delta
\right)^{-\f{n_\perp}{2}} 
\eqx 
where $\det \Delta$ is the determinant of the Laplacian operator, with
Dirichlet boundary conditions, on
the rectangle defined by the range of variation of $\rho$ and $\sg$.
$D$ is the number of massless modes (typically the dimension of space
time in which the string theory lives i.e. 10 for the superstring) 
while the $-2$ stands for the
contribution of the ghosts. In the AdS black hole background, the
effective number of transverse massless modes is
$n_\perp=10-2-1=7$. The additional subtraction of $-1$ comes from the
fact that due to the curved character of the metric one bosonic mode
becomes massive \cite{kinar}. For ordinary superstring theory in flat 10
dimensions, we would have to include also the contribution of
worldsheet fermions which would exactly cancel the bosonic one. However, as
was argued in \cite{kinar}, due to the appearance of a nonvanishing
Ramond-Ramond background field in the
AdS black hole geometry, all the fermions become massive and thus give
subleading contributions. 

The determinant of the Laplacian in (\ref{e.detform}) has to be
calculated for a rectangle of size $a\times b$ where
\eqn
a &=& L\\
b &=& \f{2L}{\th} \log \left( pT+\sqrt{1+p^2 T} \right)
\eqnx 
This determinant can be evaluated using $\zeta$ function regularization
in a calculation equivalent to the L\"uscher term
computation (c.f. (\ref{e.fltr})), and for high energies we obtain
\cite{fluct}   
\eq
\label{e.fluct}
Fluctuations=
\exp \left( n_\perp \cdot
\f{\pi}{24} \cdot \f{a}{b} \right)=
\exp \left( n_\perp \cdot \f{\pi}{24} \cdot
\f{\th}{2 \log\left( pT +\sqrt{1+p^2T^2} \right)} \right)
\ .
\eqx
Here we kept the piece which dominates after continuation to
Minkowski space (then $a/b ={\cal O}(\log s) \gg 1$).

We now have to perform the same analytical continuation to Minkowski
space as we did for the area in the preceding section. Namely we let
$log \to log -2\pi i n$. Furthermore we neglect the logarithmic $T$
dependent terms. The outcome is
\eq
\label{e.fluctsp}
Fluctuations=e^{\f{n_\perp}{96}\log s}=s^{\f{n_\perp}{96}}
\eqx

Putting together the above result and the contribution obtained in the
previous section we obtain finally for the trajectory
\eq
(prefactors)\cdot  s^{1+\f{n_\perp}{96}+\f{\alef}{4} t}
\eqx
The values for $n_\perp$ suggested by the AdS/CFT correspondence would
give an intercept of $1.073$ (or 1.083 for $n_\perp=8$), very close to
the observed soft Pomeron intercept of 1.08. The phenomenological
value of $\alef$ extracted from the static quark-antiquark potential
is $\alef\sim 0.9 \, GeV^{-2}$. This gives the slope $0.225$ in
comparison with the observed one of $0.25$.

\section{Reggeon exchange}

It is interesting to see, whether in the same framework one can obtain
an analogous description of Reggeon exchange, which should dominate
for processes {\em with} an exchange of quantum numbers in the $t$
channel. The problem is interesting as phenomenologically Reggeon
trajectories are quite different from the one of the Pomeron. The
slope is four times larger, while the dominant intercepts are around
0.5 (they depend on the trajectory) instead of 1.08. The Reggeon
amplitudes typically behave like $s^{0.5+\alef t}$.

\begin{figure}[th]
\centerline{\epsfxsize=8cm \epsfbox{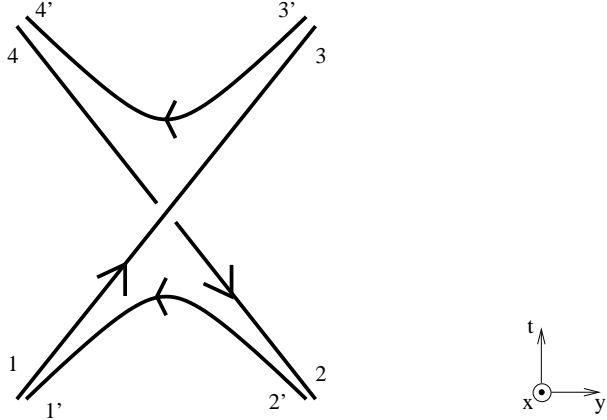}}
\caption{Spacetime picture of a meson-meson scattering process  mediated by
Reggeon exchange. The impact parameter axis is perpendicular to the
longitudinal $t-y$ plane.}
\end{figure}

We will now briefly relate the relevant results of \cite{usr}.
Since Reggeon exchange always involves an exchange of quarks in the
$t$ channel (see the schematic spacetime picture of the process in
fig. 3), the eikonal approximation which allowed to express the
scattering amplitude as a correlation function of Wilson {\em lines},
is no longer valid. The approach used in \cite{usr} uses the worldline
formalism which expresses the fermionic propagator in an external
gauge field $\AAA$ as a path integral over quark trajectories
\cite{Polsf,Korsf1,Korsf2,NRZ}: 
\eq
\label{e.wlprop}
S(x,y | \AAA)=\int \DD x^\mu(\tau) \, e^{-m \cdot Length} \cdot \left\{
\mbox{\rm Spin Factor} \right\} \cdot e^{i\int_{trajectory} A}
\eqx  
where the $\left\{\mbox{\rm Spin Factor} \right\}$ keeps track of the
spin 1/2 nature of the quarks. In the above expression the colour and
spin parts do 
factorize, which is very convenient for calculations using various
models of the nonperturbative gluonic vacuum.

In \cite{usr} we used the expressions (\ref{e.wlprop}) for the exchanged
quarks and eikonal approximation for the spectator quarks. The gauge
field dependent part became a Wilson loop expectation value, which 
was calculated using the AdS/CFT correspondence. The result again
involved the helicoid (assuming that the exchanged quarks were light).
At this stage one obtains an effective action for the trajectories of
the exchanged quarks.
Finally we performed this remaining path integral over the trajectories
of the exchanged quarks (now constrained to lie on the helicoid) by
saturating it with the classical saddle point solution.

The resulting saddle point was imaginary leading to (i) an inelastic
amplitude, and (ii) a linear Regge slope $\alef t$. The contribution of the
spin factors gave a $1/s$ suppression so at this stage the energy
dependence of the Reggeon exchange amplitude was
\eq
s^{0+\alef t}
\eqx 
When we include the contribution of the fluctuations of the string
worldsheet around the helicoid, and continue analytically this
expression to the saddle point configuration we obtain a shift of the
intercept by $n_\perp/24$. The final result is thus:
\eq
s^{0+\f{n_\perp}{24}+\alef t}
\eqx
We note that within this framework, we obtain the factor of 4 between
the slopes of the Reggeon and the `Pomeron'-like trajectories. The
intercept of the Reggeon is also four times larger than the
difference $\al_{Pomeron}(0)-1$. 

\section{Discussion}

In this paper we described the application of the AdS/CFT
correspondence to the study of high energy scattering amplitudes in
the Regge limit. In the eikonal approximation these amplitudes were
reduced to the evaluation of a Wilson loop correlation function. Using
the AdS/CFT correspondence for a confining theory this reduced the
problem to studying properties of minimal surfaces of the helicoidal
type. The linear slope of the Regge trajectory arose when making
analytical continuation from Euclidean to Minkowski space \cite{us2}, while a
shift of the intercept appeared after including the contibution of
quadratic fluctuations of the string worldsheet around the helicoid
\cite{fluct}. The resulting trajectory is very close to the
experimentally observed soft Pomeron trajectory. 

In the above approach it was crucial that we
considered a theory with confinement, which in the setting of the
black hole background allows one to use a flat space approximation for
solving the minimal surface problems. For smaller impact parameters,
where confinement is not important the resulting formulas would be
quite different and would also depend on the specific type of gauge
theory considered (which would translate to different properties of
the metric near the boundary).

Finally we briefly reported on the work of \cite{usr} where amplitudes
with Reggeon exchange were considered. These scattering processes
involve an exchange of a quark-antiquark pair in the $t$ channel and
thus necessarily require going beyond the eikonal approximation. In
order to overcome this difficulty we used the worldline formalism which
expresses (Euclidean) fermionic propagators in a background gauge
field as a path integral over quark trajectories. The averaging over
gauge fields was done using again AdS/CFT correspondence and
helicoidal minimal surfaces. The remaining path integral was done by
saddle point leading to an inelastic amplitude with a linear Regge
trajectory with {\em four} times larger slope than the one for the
Pomeron. Fluctuations led to an intercept of the order of 0.33.
This intercept is lower than the dominant intercepts observed
experimentally. A possible reason for this discrepancy was our
assumption that the exchanged quark trajectories do not deform the
helicoid minimal surface. In addition we did not evaluate fluctuations
of the boundaries of the helicoid. Due to high nonlinearity it is
difficult to estimate whether this may or may not modify the intercept.
   
It seems that the framework of AdS/CFT correspondence is very convenient to
study the interplay of soft scattering amplitudes and confinement. The
familiar Regge trajectories arise here in a nonstandard manner through
the properties of the helicoid minimal surface. In addition two types
of Regge trajectories appear naturally, corresponding qualitatively
(and partly quantitatively) to the observed Pomeron and Reggeon trajectories. 
It would be interesting to address a number of further problems in
this framework. In particular the transition to smaller impact
parameters which would necessarily involve a departure from the flat metric
approximation used here. Another very interesting but difficult problem
would be to understand how unitarization occurs within the framework
of the AdS/CFT correspondence.

\bigskip
\bigskip

\noindent{\bf Acknowledgments.} The vast majority of the results
reported here were obtained in collaboration with Robi Peschanski.
This work was supported in part by KBN grant 2P03B01917.


\begin{thebibliography}{99}

\bibitem{la92} \rr{A. Donnachie and P.V. Landshoff}{Total cross 
sections, Phys.Lett.}{B296}{(1992) 227};\\ 
\rr{A. Donnachie and P.V. Landshoff}{Soft 
interactions,}{}{hep-ph/9703366}. 

\bibitem{us1}
\rr{R.A. Janik and R. Peschanski}{High energy scattering and the
AdS/CFT correspondence, Nucl. Phys.}{B565}{(2000) 193, hep-th/9907177}; \\
\rr{R.A. Janik}{Gauge Theory Scattering from the
AdS/CFT correspondence, Cargese summer school 1999,}{}{hep-th/9909124}. 

\bibitem{us2}
\rr{R.A. Janik and R. Peschanski}
{Minimal surfaces and Reggeization in the AdS/CFT
correspondence, Nucl. Phys.}{B586}{(2000) 163, hep-th/0003059}; \\
\rr{R. Peschanski}{High Energy Scattering from the $AdS/CFT$
Correspondence, Invited talk at ``DIS 2000'', 25-30 April 2000,
Liverpool,}{}{hep-ph/0006243}.

\bibitem{fluct}
\rr{R.A. Janik}{String Fluctuations, AdS/CFT and the Soft Pomeron
Intercept, Phys. Lett.}{B500}{(2001) 118, hep-th/0010069}.


\bibitem{usr}
\rr{R.A. Janik and R. Peschanski}{Reggeon exchange from
AdS/CFT,}{}{hep-th/0110024}. 

\bibitem{SRZ} \rr{M. Rho, S.-J. Sin and I. Zahed}{Elastic
Parton-Parton Scattering from AdS/CFT, Phys. Lett.}{B466}{(1999) 199,
hep-th/9907126}. 

\bibitem{ma98} \rr{J. Maldacena}{The Large N Limit of Superconformal
Field Theories and Supergravity, Adv. Theor. Math. Phys.}{2}{(1998)
231, hep-th/9711200}; \\ 
\rr{S.S. Gubser, I.R. Klebanov and
A.M. Polyakov}{Gauge Theory Correlators from Non-Critical String
Theory, Phys. Lett.}{B428}{(1998) 105, hep-th/9802109}; \\
\rr{E. Witten}{Anti De Sitter Space And Holography,
Adv. Theor. Math. Phys.}{2}{(1998) 253, hep-th/9802150}.

\bibitem{ma99} \rr{O. Aharony, S.S. Gubser, J. Maldacena, H. Ooguri
and Y. Oz}{Large $N$ field theories, String Theory and
Gravity, Phys.Rept.}{323}{(2000) 183, hep-th/9905111}.

\bibitem{SZN}
\rr{E.~V.~Shuryak and I.~Zahed}{Instanton-induced effects in QCD
high-energy scattering, Phys.\ Rev.}{D62}{(2000) 085014,
hep-ph/0005152}; \\
\rr{M.~A.~Nowak, E.~V.~Shuryak and I.~Zahed}{Instanton-induced
inelastic collisions in QCD, Phys.\ Rev.}{D64}{(2001) 034008,
hep-ph/0012232}.

\bibitem{KKL}
\rr{D.~E.~Kharzeev, Y.~V.~Kovchegov and E.~Levin}{QCD instantons and
the soft Pomeron, Nucl.\ Phys.}{A690}{(2001) 621,
hep-ph/0007182}.

\bibitem{Nacht} \rr{O. Nachtmann}{Considerations concerning
diffraction scattering in quantum chromodynamics,
Ann. Phys. (NY)}{209}{(1991) 436}.
 
\bibitem{Nachtr} For a review see \rr{O. Nachtmann}{High
Energy Collisions and Nonperturbative QCD,}{}{hep-ph/9609365}.

\bibitem{BGKR} \rr{M.~Bianchi, M.~B.~Green, S.~Kovacs and
G.~Rossi}{Instantons in supersymmetric Yang-Mills and D-instantons in
IIB  superstring theory, JHEP}{9808}{(1998) 013, hep-th/9807033}.

\bibitem{loopsads} 
\rr{J. Maldacena}{Wilson loops in large N field theories,
Phys. Rev. Lett.}{80}{(1998) 4859, hep-th/9803002}; \\
\rr{S.-J. Rey and J. Yee}{Macroscopic strings as heavy quarks in large
$N$ gauge theory and anti-de Sitter supergravity,}{}{hep-th/9803001}; \\
\rr{J. Sonnenschein and A. Loewy}{On the Supergravity Evaluation of
Wilson Loop Correlators in Confining Theories, JHEP}{0001}{(2000) 042,
hep-th/9911172}. 

\bibitem{DG} \rr{N.~Drukker, D.~J.~Gross and H.~Ooguri}{Wilson loops
and minimal surfaces, Phys. Rev.}{D60}{(1999) 125006, hep-th/9904191}.

\bibitem{wittenbh}
\rr{E.~Witten}{Anti-de Sitter space, thermal phase transition, and
confinement in  gauge theories, Adv.\ Theor.\ Math.\ Phys.}{2}{(1998) 505,
hep-th/9803131}.

\bibitem{luscher} \rr{M. L\"uscher, K. Symanzik and
P. Weisz}{Anomalies Of The Free Loop Wave Equation In The Wkb
Approximation, Nucl.Phys.}{B173}{(1980) 365}. 
\bibitem{alvarez} \rr{O. Alvarez}{The Static Potential In String Models,
Phys.Rev.}{D24}{(1981) 440}. 
\bibitem{frts} \rr{E.S. Fradkin and A.A. Tseytlin}{On Quantized String Models,
Ann. Phys. (NY)}{143}{(1982) 413}. 
\bibitem{arvis} \rr{J.F. Arvis}{The Exact Q Anti-Q Potential In Nambu
String Theory, Phys.Lett.}{B127}{(1983) 106}. 

\bibitem{kinar} \rr{Y. Kinar, E. Schreiber, J. Sonnenschein and
N. Weiss}{Quantum fluctuations of Wilson loops from string
models, Nucl. Phys.}{B583}{(2000) 76, hep-th/9911123}. 

\bibitem{Megg} \rr{E. Meggiolaro}{The high--energy quark--quark
scattering: from Minkowskian to Euclidean theory,
Z. Phys.}{C76}{(1997) 523, hep-th/9602104};\\
\rr{E. Meggiolaro}{The analytic continuation of the high-energy quark-quark
scattering amplitude, Eur. Phys. J.}{C4}{(1998) 101, hep-th/9702186};\\
\rr{E. Meggiolaro}{A remark on the high--energy quark--quark scattering and the
eikonal approximation, Phys. Rev.}{D53}{(1996) 3835, hep-th/9506043};\\
\rr{E. Meggiolaro}{The analytic continuation of the high-energy parton-parton 
scattering amplitude with an IR cutoff,}{}{hep-ph/0110069}.

\bibitem{Polsf}
\rr{A.~M.~Polyakov}{Fermi-Bose Transmutations Induced By Gauge Fields,
Mod.\ Phys.\ Lett.}{A3}{(1988) 325}.

\bibitem{Korsf1}
\rr{G.~P.~Korchemsky}{Quantum geometry of Dirac fermions, 
Int.\ J.\ Mod.\ Phys.}{A7}{(1992) 339}.

\bibitem{Korsf2}
\rr{G.~P.~Korchemsky}{Quantum Geometry Of The Dirac Fermions,
Phys.\ Lett.}{B232}{(1989) 334}.

\bibitem{NRZ}
\rr{M.~A.~Nowak, M.~Rho and I.~Zahed}{Spin factors and geometric
phases in arbitrary dimensions, Phys.\ Lett.}{B254}{(1991) 94}.


\end{thebibliography}
\end{document}